\documentclass[aps,showpacs,showkeys,twocolumn]{revtex4}
\usepackage{amsmath,amsthm,amssymb,epsfig,alltt}

\begin{document}

\def\a{\alpha}
\def\b{\beta}
\def\bt{{\beta_T}}
\def\l{\lambda}
\def\e{\epsilon}
\def\bole{{\boldsymbol \epsilon}}
\def\p{\partial}
\def\m{\mu}
\def\n{\nu}
\def\t{\tau}
\def\th{\theta}
\def\bolth{{\boldsymbol \theta}}
\def\s{\sigma}
\def\g{\gamma}
\def\o{\omega}
\def\r{\rho}
\def\half{\frac{1}{2}}
\def\hatt{{\hat t}}
\def\hatx{{\hat x}}
\def\hatp{{\hat p}}
\def\hatX{{\hat X}}
\def\hatY{{\hat Y}}
\def\hatP{{\hat P}}
\def\haty{{\hat y}}
\def\hatth{{\hat \theta}}
\def\hatta{{\hat \tau}}
\def\hatrh{{\hat \rho}}
\def\hatva{{\hat \varphi}}
\def\barx{{\bar x}}
\def\bary{{\bar y}}
\def\barz{{\bar z}}
\def\baro{{\bar \omega}}
\def\sp{\sigma^\prime}
\def\nn{\nonumber}
\def\cb{{\cal B}}
\def\2pap{2\pi\alpha^\prime}
\def\wideA{\widehat{A}}
\def\wideF{\widehat{F}}
\def\beq{\begin{eqnarray}}
 \def\eeq{\end{eqnarray}}
 \def\4pap{4\pi\a^\prime}
 \def\xp{{x^\prime}}
 \def\sp{{\s^\prime}}
 \def\ap{{\a^\prime}}
 \def\tp{{\t^\prime}}
 \def\zp{{z^\prime}}
 \def\xpp{x^{\prime\prime}}
 \def\xppp{x^{\prime\prime\prime}}
 \def\barxp{{\bar x}^\prime}
 \def\barxpp{{\bar x}^{\prime\prime}}
 \def\barxppp{{\bar x}^{\prime\prime\prime}}
 \def\barchi{{\bar \chi}}
 \def\baro{{\bar \omega}}
 \def\bpsi{{\bar \psi}}
 \def\barg{{\bar g}}
 \def\barz{{\bar z}}
 \def\bareta{{\bar \eta}}
 \def\ta{{\tilde \a}}
 \def\tb{{\tilde \b}}
 \def\tc{{\tilde c}}
 \def\tz{{\tilde z}}
 \def\tJ{{\tilde J}}
 \def\tpsi{\tilde{\psi}}
 \def\tal{{\tilde \alpha}}
 \def\tbe{{\tilde \beta}}
 \def\tga{{\tilde \gamma}}
 \def\tchi{{\tilde{\chi}}}
 \def\barth{{\bar \theta}}
 \def\bareta{{\bar \eta}}
 \def\barom{{\bar \omega}}
 \def\bomega{\boldsymbol \omega}
\setcounter{page}{1}
\title[]{Canonical Quantization of \\The Dissipative Hofstadter Model
}

\author{Taejin Lee}
\affiliation{
Department of Physics, Kangwon National University, Chuncheon 200-701
Korea}

\email{taejin@kangwon.ac.kr}

\begin{abstract}
We perform canonical quantization of the dissipative Hofstadter model, 
which has a wide range of applications in condensed matter physics and string theory. 
The target space duality and the non-commutative algebra developed in string theory are 
discussed for the model. We show that the target space duality transformation of closed string theory, $O(2,2;R)$, removes the interaction with a uniform magnetic field. 
The $O(2,2;R)$ dual transformation changes the 
basis of oscillator operators so that the algebra of the string coordinate operators at the boundary 
become non-commutative. In the zero temperature limit, the non-commutative algebra of open string 
theory emerges. We also developed the boundary state formulation for the dissipative Hofstadter model.
\end{abstract}




\maketitle

\begin{center}
{\bf I. Introduction}
\end{center}

String theory proves to be an imperative tool to explore the critical behavior of the quantum dissipative system at nano-scale. The dissipative Hofstadter model \cite{dwahm}
describes the quantum particles moving in two dimensions, subject to a uniform magnetic field, a periodic potential and a dissipative force.
Caldeira and Leggett \cite{caldeira83ann,caldeira83phy} introduced a bath or environment which consists 
of an infinite number of harmonic oscillators and coupled it to the quantum particle system. 
In the classical limit the coupling of the system to the environment produces the dissipative force. 
Integrating out the bath degrees of freedom in the action results in a non-local effective interaction in 
quantum theory. Unlike one dimensional theories with local interactions only, the quantum dissipative models
with the non-local interactions exhibit rich critical behavours; phase transitions. 
The non-trivial critical behavour of the 
quantum dissipative model has been observed first by Schmid \cite{schmid,fisher;1985,kane;1992b} 
in a simpler model, now called the Schmid model,
which has the periodic potential only. Depending on the value of frictional constant, the Schmid model
has two phases; the delocalized phase where the periodic potential is weak and the localizd phase where
the periodic potential is strong. In the localized phase the particles are mostly localized in the 
minima of the potential. 

If a uniform mangetic field is turned on, the Schmid model becomes the dissipative Hofstadter model,
of which phase diagram has a more complex and interesting structure. 
The dissipative Hofstadter model has been studied extensively for more than two decades, 
since it is known to have a wide range of applications in condensed matter physics such as  
the Josephson junction arrays \cite{larkin,fazio,sodano},
the Kondo problem \cite{Affleck:1990by}, the study of 
one-dimensional conductors \cite{Kane:1992a}, 
tunnelling between Hall edge states \cite{kane2}, 
and junctions of quantum wires~\cite{Oshikawa:2005fh}. 
However, our understanding of the model is still far from complete yet. 
We need a concrete and complete analysis to resolve many important issues associated with 
the model. As a step toward this goal, we perform a canoncal quantization of the model in this 
paper. The dissipative quantum models can be reformulated as string theories on a disk \cite{callan90}.
The time dimension of the model is mapped onto the boundary of the disk diagram and the 
the bulk degrees of freedom of the string play the role of the bath to 
produce the non-local dissipative term at the boundary. The non-local dissipative term in 
the one dimensional model is traded with the local Polyakov term in two dimensional string 
theory. The interaction term with the magnetic field corresponds to the interaction term with
the Neveu-Schwarz (NS) B field on the two dimensional worldsheet in string theory.
The advantage of this approach is clear in that we only have to deal with a local two-dimensional 
quantum field theory where string theory techniques are readily available.




\begin{center}
{\bf II. String Theory Action and Boundary State}
\end{center}

The non-local action for the dissipative Hofstadter model is given as follows
\beq 
S &=& \frac{\eta}{4\pi \hbar} \int^{\bt/2}_{-\bt/2} dt dt^\prime 
\frac{\left(X^I(t) - X^I(t^\prime)\right)^2}{(t-t^\prime)^2}  \nn\\
&& +\frac{ieB_H}{2\hbar c} \int^{\bt/2}_{-\bt/2} dt \epsilon^{IJ}
\p_t X^I X^J \nn\\
&& + \frac{V_0}{\hbar} \int^{\bt/2}_{-\bt/2} dt \sum_I
\cos \frac{2\pi X^I}{a},
\eeq
where $\eta$ is the friction coefficient and $I,J = 1, 2$. The first term is the non-local effective action
which is obtained by integrating out the bath degrees of freedom. The second term is the interaction term with
the uniform magnetic field $B_H$ and the third term corresponds to the periodic potential due to the one dimensional lattice. 
The zero temperature limit is achieved by taking the limit where $\bt \rightarrow 
\infty$.   
As in the case of the Schmid model, the action for the dissipative Hofstadter model can be interpreted 
as the boundary effective action 
for the string with a periodic tachyon potential and the Neveu-Schwarz (NS) B field with a boundary condition $X^I(\t=0,\s) = X^I(\s)$,
\beq
S&=&  \frac{1}{4\pi}
\int d\t d\s E_{IJ} \left(\p_\t - \p_\s\right) X^I 
\left(\p_\t + \p_\s\right) X^J \nn\\
&&- \frac{V}{2} \int d\s \sum_I \left(e^{iX^I}+ 
e^{-iX^I}\right)
\eeq 
where $E_{IJ} = \a\delta_{IJ} + 2\pi B_{IJ}= 
\a\delta_{IJ} + \b \epsilon_{IJ}$, $\a = 1/\ap$, $\s = 2\pi t/\bt$,
and $2\pi\b = \frac{eB_H}{\hbar c}a^2$. The $2 \times 2$ matrix $E$ is called the background matrix.

Let us turn off the periodic potential for the time being. Then the action reduces to the
closed string action in the presence of the NS B-field
\beq \label{bulk}
S_E &=& \frac{1}{\4pap} \int d^2\xi \Biggl[
\frac{\p X^I}{\p \xi^\a} \frac{\p X^I}{\p \xi^\a} \nn\\
&&~~~~~~~~~~~~~~
+ \2pap B_{IJ} \e^{\a\b} \frac{\p X^I}{\p \xi^\a} \frac{\p X^J}{\p \xi^\b} 
\Biggr] \\
&=& \frac{1}{4\pi} \int d^2\xi E_{IJ} (\p_\t - \p_\s)X^I
(\p_\t + \p_\s) X^J,\nn
\eeq
where $(\xi^1,\xi^2)= (\t,\s)$.
If we interchange the worldsheet parameters, $\t \rightarrow \s$, $\s \rightarrow \t$, 
we have the open string action, which has been discussed extensively in the literature 
in connection with the non-commutative geometry of the D-branes. 

Taking advantage of string theory and the boundary formulation, 
we may evaluate efficiently the partition function and the correlation functions of 
operators ${\cal O}(t_i), \, i=1, \dots, n$ as follws
\beq
Z =\int D[X] \exp \left(-S\right) = \langle 0 \vert B \rangle, \nn\\
\langle {\cal O} (t_1) \cdots {\cal O}(t_n) \rangle 
= \langle 0 \vert \langle {\cal O} (\s_1) \cdots {\cal O}(\s_n)\vert B \rangle
\eeq
where $\s_i = \frac{2\pi t_i}{\bt}$.
The boundary state $\vert B\rangle$ is formally written as
\beq \label{bound1}
\vert B\rangle = \exp \left[ \frac{V}{2} \int_M d\s \sum_I \left(e^{iX^I} + e^{-iX^I} \right)
\right] \vert B_E \rangle.
\eeq
The boundary condition for the boundary state $|B_E\rangle$ is determined by the bulk action 
Eq.(\ref{bulk}). 
Since the interaction with the magnetic field can be also writtens as a boundary term 
\beq
\frac{1}{2}\int_M d\t d\s B_{IJ} \e^{\a\b} \frac{\p X^I}{\p \xi^\a} \frac{\p X^J}{\p \xi^\b} = 
\frac{1}{2}\int_{\p M} B_{IJ} X^I \frac{X^J}{\p \s},
\eeq
we may write the boundary state $\vert B_E \rangle$ as
\beq \label{bound2}
\vert B_E \rangle = \exp\left[\frac{1}{2}\int_{\p M} B_{IJ} X^I \frac{X^J}{\p \s} \right]
\vert N \rangle
\eeq
where $\vert N\rangle$ is the Neumann boundary state. 

If we keep the Polyakov action only 
as a bulk action and treat both magnetic interation term and periodic potential term as 
boundary interaction terms, the quantization procedure is same as the standard one: The 
Lagrangian is just the Polyakov term
\beq
L = \frac{1}{\4pap} \frac{\p X^I}{\p \xi^\a} \frac{\p X^I}{\p \xi^\a} .
\eeq
The canonical momenta and the Hamiltonian are given
\begin{subequations} 
\beq
P^I &=& \frac{1}{\2pap}\p_\t X^I, \\
H &=& \int d\s \half\left[ \2pap P^I P^I +\frac{1}{\2pap}
\p_\s X^I \p_\s X^I \right]. 
\eeq
\end{subequations}
Expanding the canonical variables in terms of normal modes, we have
\begin{subequations} 
\beq
X^I(\s) &=& \omega^I \s + \sum_n X^I_n e^{in\s}, \\ 
P^I(\s) &=& \sum_n P^I_n e^{-in\s},
\eeq
\end{subequations} 
where $\o^I$, $I=1,2$ are winding modes. If we rewrite the Hamiltonian in terms 
of normal modes, 
\beq
H/(2\pi) &=& \frac{1}{2}\left[(\2pap) \left(\frac{p^I}{2\pi}\right)^2+  \frac{1}{2\pi\ap}(\o^I)^2 \right]\nn\\
&&
+ \sum_{n=1} \left[ \2pap P^I_{-n} P^I_{n}  + \frac{n^2}{\2pap} X^I_n X^I_{-n}\right]
\eeq
where $p^I = 2\pi P^I_0$. 
Since we have a periodic potential $e^{\pm X^I}$, the action should be defined on a torus;
$X^I \sim X^I + 2\pi$, $I=1, 2$. Accordingly the winding modes and the zero modes of the momenta 
have integer values
\beq
\o^I = n^I, \quad p^I= m^I, \quad n^I~{\rm and}~m^I~{\rm are}~{\rm integers}.
\eeq 
It is more convenient to rewrite the canonical variables and Hamiltonian in terms of oscillator 
operators
\begin{subequations} 
\beq
X^I(\s,\t) &=& x^I + \o^I \s + \ap p^I \t  \nn\\
&&+i\sqrt{\frac{\ap}{2}} \sum_{n\not=0} \frac{1}{n}
\Bigl[\a^I_n e^{-in(\t-\s)} \nn\\
&&+ \ta^I_n e^{-in(\t+\s)}\Bigr], \label{oscillator}\\
2\pi P^I (\s,\t) &=& p^I + \sqrt{\frac{1}{2\ap}}, \sum_{n\not=0}
\Bigl[\a^I_n e^{-in(\t-\s)} \nn\\
&&+ \ta^I_n e^{-in(\t+\s)}\Bigr],\\
H &=& \frac{1}{2}\left[\ap (p^I)^2 + \frac{1}{\ap} (\o^I)^2 \right] \nn\\
&&+ \sum_{n=1} \a^I_{-n} \a^I_n + \sum_{n=1} \ta^I_{-n} \ta^I_n
\eeq
\end{subequations} 
where the fundamental commutation relations are
\beq
\left[ x^I,p^J\right]&=&i\delta^{IJ},
 ~ \left[ \a^I_m, \a^J_n \right] = \delta^{IJ}m\delta_{m+n},\nn\\ 
\left[\tilde\a^I_m, \tilde\a^J_n \right] &=&\delta^{IJ} m\delta_{m+n}.
\eeq
It looks straightforward to evaluate the partition function by using Eqs.(\ref{bound1},\ref{bound2}) 
\begin{subequations} 
\beq
\vert B\rangle &=& \exp \left[ \frac{V}{2} \int_M d\s \sum_I \left(e^{iX^I} + e^{-iX^I} \right)
\right] \nn\\
&&\exp\left[\frac{1}{2}\int_{\p M} B_{IJ} X^I \frac{X^J}{\p \s} \right]
\vert N \rangle,\\
|N \rangle &=& \prod_{I=1,\, n=1} e^{-\frac{1}{n} \a^I_{-n} \ta^I_{-n}} \vert 0 \rangle,
\nn\\
\a^I_n \vert N\rangle &=& -\ta^I_{-n} \vert N\rangle,~~  p^I\vert N\rangle =0, \quad n>0 .
\eeq
\end{subequations} 
However, it turns out to be a formidable task to evaluate the partition function or the correlation 
functions explicitly in this way, especially when the potential or the interaction with the magnetic
field are strong. We should look for a better way to quantize the system, which renders 
the explicit evaluations feasible.

\begin{center}
{\bf III. Canonical Quantization and Noncommutativity}
\end{center} 

If we keep the Poyakov term only as a bulk term, the quantization procedure is just the standard one,
which can be found in the literature. But explicit evaluations of the partition function and the
correlation functions are very difficult. If the periodic potential and the magnetic interaction 
term are not weak, the formal expressions of the boundary state and the partion function are not
useful. As in the case of open string, moving in the NS B-field background, we have an alternative
choice to quantize the system: The magnetic interaction term is quadratic in coordinate variables.
So we may include the magnetic interaction term as a part of the kinetic bulk action,
treating the periodic potential as an only interaction term.

The classical Lagrangian, which we will quantize, consists of two terms; the
Polyakov term and the magnetic interaction term,
\beq \label{class}
L = \frac{1}{4\pi} \left[
\frac{\p X^I}{\p \xi^\a} g_{IJ} \frac{\p X^J}{\p \xi^\a} 
+ 2\pi B_{IJ} \e^{\a\b} \frac{\p X^I}{\p \xi^\a} \frac{\p X^J}{\p \xi^\b} 
\right].
\eeq
The canonical momenta and the Hamiltonian are given by
\begin{subequations} 
\beq
P^I &=& \frac{1}{\2pap } \left[\p_\t X^I + \2pap B^{IJ} \p_\s X^J \right],\label{momenta} \\
H &=& \half \int d\s \Biggl[ \2pap \left(P^I - B^{IJ} \p_\s X^J\right)^2\nn\\
&&~~~~~~~~~~~~~~~ + \frac{1}{\2pap}
\p_\s X^I \p_\s X^I \Biggr].\label{hamiltonian}
\eeq
\end{subequations} 
where $g_{IJ} =\a \delta_{IJ}$ is the target space metric (the symmetric part of $E$).
Since the magnetic interaction term is linear in 
$\p_\t X^I$, the canonical momenta are shifted by $B^{IJ}\p_\s X^J$.
If we expand the canonical string variables in terms of normal 
modes, as before
\begin{subequations}  
\beq 
X^I(\s) &=& \omega^I \s + \sum_n X^I_n e^{in\s}, \\ 
P^I(\s) &=& \sum_n P^I_n e^{-in\s},
\eeq
\end{subequations} 
we have 
\beq
H/(2\pi) &=& \frac{1}{2}\left[(\2pap) (\frac{p^I}{2\pi} - B^{IJ}\o^J)^2 + \frac{1}{2\pi\ap} (\o^I)^2\right] \nn\\
&& \quad + \sum_{n=1} \Biggl[ \2pap (P^I_{-n} - i n B^{IJ} X^J_n)\nn\\
&&(P^I_{n} +i n B^{IK} X^K_{-n}) + \frac{n^2}{\2pap} X^I_n X^I_{-n}\Biggr].
\eeq
Since the magnetic interaction term shifts the momenta variables only, the coordinate variables may 
be written in terms of the oscillator operators as before Eq.(\ref{oscillator}) at $\t=0$
\beq
X^I(\s,0) &=& x^I + \o^I \s  \nn\\
&&+i\sqrt{\frac{1}{2}} \sum_{n\not=0} \frac{1}{n}
\left[\a^I_n e^{in\s} + \ta^I_n e^{-in\s}\right].
\eeq
The spectrum of the Hamiltonian for non-zero modes would not be affected by the magnetic interaction.
The coordinate variables at $\t$ follows from the time evolution equation: 
${\cal O}(\s,\t) = e^{i\t H} {\cal O} e^{-i\t H}$,
\beq
X^I(\s,\t) &=& x^I + \o^I \s + \left(\ap p^I - 2\pi\ap B^{IJ} \o^J\right)\t \nn\\
&& + 
i\sqrt{\frac{1}{2}} \sum_{n\not=0} \frac{1}{n}\Bigl[\a^I_n e^{-in(\t-\s)}\nn\\
&& + \ta^I_n e^{-in(\t+\s)}\Bigr]. 
\eeq
Then the expression of the momentm variables and the Hamiltonian in term of the oscillator operators follows from Eqs.(\ref{momenta},\ref{hamiltonian}), 
\begin{subequations} 
\beq
2\pi P_I (\s,\t) &=& p_I + \sqrt{\frac{1}{2}} \sum_{n\not=0}
\Bigl[E_{IJ} \a^J_n e^{-in(\t+\s)} \nn\\
&&+ E_{IJ}^T \ta^J_n 
e^{-in(\t-\s)}\Bigr], \\
H &=& \half {\bf Z}^T {\bf M}(E){\bf Z} +\sum_{n=1} \a^I_{-n} g_{IJ} \a^J_n \nn\\
&& + \sum_{n=1} \ta^I_{-n}g_{IJ}\ta^J_n.
\eeq
\end{subequations} 
where $g_{IJ} = \a \delta_{IJ}$, ${\bf Z}^T = (\o^1, \o^2, p_1, p_2)$ and
\beq \label{me}
{\bf M}(E) &=& \left(\begin{array}{cc}
  \ap^{-1}\left(I-(\2pap {\bf B})^2\right) & \2pap  {\bf B} \\
 - \2pap  {\bf B} & \ap I  
\end{array} \right) \nn\\
&=& \left(\begin{array}{cc}
  \left(\frac{\a^2+\b^2}{\a}\right) I & \frac{\b}{\a} {\bole} \\
-\frac{\b}{\a} {\bole} & \frac{1}{\a} I  
\end{array} \right),
\eeq
where $I$ is the $2 \times 2$ identity matrix and $(\bole)_{IJ} = \e_{IJ}$.
The oscillator operators satisfy
\beq
\left[ x^I,p^J\right]&=&i\delta^{IJ},~
\left[ \a^I_m, \a^J_n \right] = g^{IJ}m\delta_{{m+n},0},\nn\\ 
\left[\tilde\a^I_m, \tilde\a^J_n \right] &=& g^{IJ} m\delta_{{m+n},0}, \quad g^{IJ} = \a^{-1}\delta^{IJ}.
\eeq

The canonical quantization of the system has not been completed yet. We should find an appropriate
representation of the boundary conditions to construct the boundary state.
The boundary conditions follows from the variation of the action for the Lagrangian 
Eq.(\ref{class})
\beq
\p_\t X^I + \2pap B^{IJ} \p_\s X^J = 0, \quad 
{\rm at}\,\,\, \t = 0.
\eeq
In terms of oscillator operators, they are written as
\beq
E_{IJ}^T \a^J_{n} + E_{IJ} \ta^J_{-n} = 0, \quad p^I = 0.
\eeq
Note that these conditions are not diagonal in the oscillator basis ${\a^I, \ta^I, I=1,2}$. We may recall
that in the canoncial quantization of the open string with the NS B-background, the Hamiltonian is diagonalized
with respect to the metric \cite{Lee;open}
\beq
G_{IJ} = (E^T g^{-1} E)_{IJ}= \left(\frac{\a^2+\b^2}{\a}\right) \delta_{IJ}.
\eeq  
We should find a new oscillator basis, in which both Hamiltonian and boundary conditions are diagonal.
To this end we apply the $O(2,2;R)$ target space duality \cite{giveon}.

It is well known that the $d$ dimensional closed string in the NS B-field background possesses the 
$O(d,d;R)$ duality \cite{giveon}. The element $T \in O(d,d;R)$ preserves the form $J$ as
\beq
T &=& \left(\begin{array}{cc} a & b \\ c & d \end{array} \right),~ 
J =\left(\begin{array}{cc} 0 & I \\ I & 0 \end{array} \right),\\
T^T J \,T &=& J ; ~ a^T c + c^T a = 0,\nn\\
b^T d+d^T b&=& 0,~ a^T d+ c^T b = I,\nn
\eeq
where $a, b, c, d$ are $d \times d$ matrices and $I$ is the identity matrix in $d$ dimensions.
For the dissipative Hofstadter model $d=2$. 
Under the $O(2,2;R)$ dual transformation the background matrix $E$ and 
the oscillator basis operators transform as
\beq
E &\rightarrow& (aE +b)(cE+d)^{-1}, \nn\\
\a_n &\rightarrow& (d-cE^T)^{-1} \b_n,\\
\ta_n &\rightarrow& (d+cE)^{-1} \tb_n, \nn  
\eeq
where $\b_n$ and $\tb_n$ are new oscillator basis operators, satisfying the following 
fundamental commutation relations
\beq
\bigl[\b^I_n, \b^J_m\bigr]&=& m G^{IJ} \delta_{n+m,0},\quad \bigl[\tb^I_n, \tb^J_m\bigr]= m G^{IJ}  \delta_{n+m,0},\nn\\
G^{IJ} &=& (G^{-1})_{IJ}.
\eeq
By some algebra, we find that the anti-symmetric part of the background matrix (the magnetic field)
is removed by the following T-dual transformation\cite{Lee;mod} generated by
\begin{subequations}   
\beq 
T &=& \left(\begin{array}{cc} I & 0 \\ \bolth/(2\pi) & I \end{array} \right);\\
&& \bolth/(2\pi) = 
\frac{1}{E} (2\pi B) \frac{1}{E^T} = \frac{\b}{\a^2+\b^2} \bole .
\eeq
\end{subequations} 
Under this T-dual transformation the oscillator basis operators and the non-zero mode parts of Hamiltonian
transform as; $\a^I_n \rightarrow g^{IJ} E_{JK} \b^K_n,\quad \ta^I_n \rightarrow g^{IJ} E^T_{JK} \tb^K_n$,
\beq
&&\sum_{n=1} \a^I_{-n} g_{IJ} \a^J_n + \sum_{n=1} \ta^I_{-n} g_{IJ} \ta^J_n \nn\\
&&~~~~~~~~~\rightarrow \sum_{n=1} \b^I_{-n} G_{IJ}  \b^J_n + \sum_{n=1} \tb^I_{-n} G_{IJ} \tb^J_n .
\eeq
The boundary conditions become the usual Neumann condition in the new oscillator basis
\beq
\b^I_{-n} + \tb^I_n = 0.
\eeq

Now we are in a position to discuss the transofrmation of the zero mode sector of the Hamiltonian.
Note that ${\bf M}(E)$ in Eq.(\ref{me}) can be written in terms of an element of $O(2,2;R)$, $g_E$ as
\begin{subequations} 
\beq
{\bf M}(E) &=& g_E (g_E)^T,\\
g_E &=& \frac{1}{\sqrt{\a}}\left(\begin{array}{cc} g & 2\pi B \\ 0 & I \end{array} \right) \,\in O(2,2;R).
\eeq
\end{subequations} 
Under the $O(2,2;R)$ dual transformation generated by $T$, $g_E$ and
${\bf M}(E)$ transform as
\begin{subequations} 
\beq
g_{E} & \rightarrow& g_{E^\prime} = T g_E \nn\\
&&~~~~~ = \sqrt{\a} 
\left(\begin{array}{cc} I & \a\b\bole \\ {\bolth}/(2\pi) & G^{-1} \end{array} \right),\\
{\bf M}(E) &\rightarrow& {\bf M}(E^\prime) = T {\bf M}(E) T^T \nn\\
&&~~~~~~~~~= g_{E^\prime} (g_{E^\prime})^T \nn\\
&&~~~~~~~~~=
\left(\begin{array}{cc} {\bf G} & 0 \\ 0 & {\bf G}^{-1} \end{array} \right),
\eeq
\end{subequations} 
where $({\bf G})_{IJ} = G_{IJ}$.
This is precisely what would be obtained when we quantize the zero mode sector of the 
closed string action with the target space metric $G_{IJ}$ without the magnetic field background. By an explicit construction of the $O(2,2;R)$ transformation, 
we have shown that the magnetic interaction term in the non-zero mode sector can be completely 
removed by a target space duality transformation. At the same time the T-dual transformation changes
the target space metric as $g_{IJ} \rightarrow G_{IJ}$. 
For the zero mode part of the Hamiltonian to be invarant under the T-dual transformation $T$, the
zero modes should transform as ${\bf Z} \rightarrow (T^T)^{-1} {\bf Z} = JTJ  {\bf Z}$, {\it i.e.}
\beq
\o^I \rightarrow \o^{\prime I}=\o^I + \frac{\th^{IJ}}{2\pi} p_J, \quad p^I \rightarrow p^I.
\eeq
It implies that for $\o^{\prime I}$ to be interpreted as winding numbers of the closed string with target space
metric $G_{IJ}$, $\o^I + \frac{\th^{IJ}}{2\pi} p_J$ must be integers. But it is possible only when
$\frac{\th^{IJ}}{2\pi}$ are integers, {\it i.e.}
\beq
\frac{\b}{\a^2+\b^2} = n, \quad n \in {\bf Z}.
\eeq
This condition defines the magic circle \cite{Lee;mod,callan91,Lee;2007}
\beq
\a^2+ \left(\b- \frac{1}{2n}\right)^2 = \left(\frac{1}{2n}\right)^2.
\eeq 
Therefore, we do not take the T-dual transfomation for the zero mode sector unless the model is not 
on the magic circle.

In the open string theory in the NS B-field background, we impose the constraints, which arise as
boundary conditions at the end points of the open string. It leads us to the open string action
with the target space metric $G_{IJ}$ but no NS B-field background. Imposing the constraints results in
the non-commutative algebra of the coordinate variables $X^I$ at the ends of the open string. 
In the closed string theory we also find that the non-commutative algebra emerges.
At the boundary, where $\t=0$, 
\beq
X^I(\s,0) &=& x^I + \o^I \s \nn\\
&& +  i\frac{1}{\sqrt{2}} \sum_{n\not=0} \frac{1}{n}
\left[\a^I_n e^{in\s} + \ta^I_n e^{-in\s}\right]. 
\eeq
Applying the T-dual transformation generated by $T$, we have
\beq \label{applying}
X^I(\s,0) &=& x^I + \o^I \s +  i\frac{1}{\sqrt{2}} \sum_{n\not=0} \frac{1}{n}
\Bigl[(g^{-1} E)^I{}_{J}\b^J_n e^{in\s} \nn\\
&&~~~+ (g^{-1}E^T)^I{}_{J} \tb^J_n e^{-in\s}\Bigr]\nn\\
&=& x^I + \o^I \s  +  i \frac{1}{\sqrt{2}}\sum_{n\not=0} \frac{1}{n}
\Biggl[\left(I + \frac{\b}{\a} \bole\right)^{IJ}\b^J_n e^{in\s} \nn\\
&&~~~+\left(I - \frac{\b}{\a} \bole\right)^{IJ}\tb^J_n e^{-in\s}\Biggr].
\eeq
Some algebra yields the commutator of the coordinate operators: 
\beq
&&\left[X^I(\s_1,0),X^J(\s_2,0)\right] \nn\\
~~~~~~~~&=& -\frac{2\b}{\a^2+\b^2} \e^{IJ} \sum_{n\not =0} \frac{1}{n} e^{in(\s_1-\s_2)} \nn\\
~~~~~~~~&=& \frac{2i}{\pi} \th^{IJ} \arctan
\left[ \frac{\sin(\s_1-\s_2)}{\cos(\s_1-\s_2) -1} \right].
\eeq
The non-commutativity becomes more transparent in the zero temperature limit; 
in the zero temperature limit we may take $\s_i = 2\pi t_i/\b_T \rightarrow 0$,
\beq
\left[X^I(\s_1,0),X^J(\s_2,0)\right] = i \th^{IJ}.
\eeq
This is precisely the non-commutative relations between the open string coordinate operators. It is
interesting to observe that the open string commutative algebra is realized in the closed string theory
in the zero temprature limit. 

The noncommutative operator algebra emerges also when we consider the operator product at the boundary,
$\t =0$. Recalling that the non-zero mode sector of the bulk action can be entirely described in terms of the coordinate variables $Z^I$ for the closed string with target space metric $G_{IJ}$, thanks to the $O(2,2;R)$ 
T-dual transformation, we define new coordinate operators at $\t=0$ as 
\beq
Z^I(\s,0) &=& x^I + \o^I \s \nn\\
&&+ i\frac{1}{\sqrt{2}} \sum_{n\not=0} \frac{1}{n}
\left[\b^I_n e^{in\s} + \tb^I_n e^{-in\s}\right]. 
\eeq
(Here the zero mode part of $Z^I$ is same as that of $X^I$.) 
But the boundary periodic potential cannot be expressed entirely in terms of $Z^I$ as we see in 
Eq.(\ref{applying}). 
The boundary state may be written as
\beq
\vert B \rangle = \exp \left[ \frac{V}{2} \int_{\p M} d\s \sum_I \left(e^{iX^I} + e^{-iX^I} \right)
\right] \vert N_E \rangle,
\eeq
where
\beq
|N_E \rangle &=& \prod_{I=1,\, n=1} e^{-\frac{1}{n} \b^I_{-n} \tb^I_{-n}} \vert 0 \rangle,\nn\\
0 &=&\left(\b^I_n + \tb^I_{-n} \right) \vert N_E\rangle.
\eeq
Expanding the boundary state in $V$, we get
\beq
\vert B \rangle &=& \sum_n \frac{1}{n!} \left(\frac{V}{2}\right)^n \nn\\
&&\prod^n_{j=1} \int d\s_j 
\exp\left[i{\bf q}_j \cdot {\bf X}(\s_j)\right]\vert B_E \rangle ,
\eeq
where $q^I_j = 0, \pm 1$. Note that since $\left(\b^J_n+\tb^J_{-n} \right) \vert B_E \rangle = 0$ and
\beq
X^I(\s,0) &=& Z^I(\s,0) \nn\\
&&+ \frac{i}{\sqrt{2}} \frac{\b}{\a} \sum_{n\not=0} \frac{1}{n}
\e^{IJ} \left(\b^J_n+\tb^J_{-n} \right) e^{in\s}, 
\eeq
we find
$\exp\left[i {\bf q} \cdot {\bf X}(\s) \right] \vert B_E \rangle =
\exp\left[i {\bf q} \cdot {\bf Z}(\s) \right] \vert B_E \rangle $.
Now consider the product of two operators
\beq
&&\exp\left[i {\bf q}_1 \cdot {\bf X}(\s_1)\right] \exp\left[i {\bf q}_2 \cdot {\bf X}(\s_2) \right]
\vert B_E \rangle \nn\\
&&~~~~~= \exp\left[i {\bf q}_1 \cdot {\bf X}(\s_1)\right] \exp\left[i {\bf q}_2 \cdot {\bf Z}(\s_2) \right]
\vert B_E \rangle.
\eeq
Making use of the identity, $e^A e^B = e^B e^A e^{[A,B]}$, we have 
\beq
&&\left[i {\bf q}_1 \cdot {\bf X}(\s_1), i {\bf q}_2 \cdot {\bf Z}(\s_2) \right] \nn\\
&=& -\frac{i}{\pi} {\bf q}_1 \cdot \bolth \cdot {\bf q}_2 \arctan\,
\left[ \frac{\sin(\s_1-\s_2)}{\cos(\s_1-\s_2) -1} \right].
\eeq
Thus, the product of the two operators acting on the boundary state can be written in the zero temperature limit as
\beq
&&e^{i {\bf q}_1 \cdot {\bf X}(\s_1)} e^{i {\bf q}_2 \cdot {\bf X}(\s_2)}
\vert B_E \rangle 
= \exp\left[-\frac{i}{2} {\bf q}_1 \cdot \bolth \cdot {\bf q}_2\right]\nn\\
&&
e^{i {\bf q}_1 \cdot {\bf Z}(\s_1)} e^{i {\bf q}_2 \cdot {\bf Z}(\s_2)}
\vert B_E \rangle .
\eeq

\begin{center}
{\bf IV. Conclusions}
\end{center}

The intimate relation between the quantum dissipative systems and the 
string theory enables us to explore the critical behaviors of the quantum dissipative 
systems. In this paper we performed the canonical quantization of the dissipative 
Hofstadter model, which has a wide range of applications in condnsed matter physics,
using the framework of the string theory and boundary state formulation.
The dissipative Hofstadter model is equivalent to the string theory with a boundary periodic 
potential in the NS B-field background on a disk. If the NS B-field or the magnetic field is 
turned off, the model reduces to the Schmid model, which corresponds to the rolling tachyon model
\cite{Sen:2002nu} which is proposed to depict the decaying unstable D-brane in string theory.  
The critical behaviours of the Schmid model and their implications for the rolling tachyon
have been discussed in refs.\cite{Lee:2005ge,Tlee:06,TLee;08}.

If we turn off the periodic potential, the dissipative Hofstadter model reduces to the Hofstadter model
\cite{dwahm}, which corresponds to the open string in the NS B-field background.   
The NS B-field, or equivalently the uniform magnetic field brings the non-commutativity into the
operator algebra of the model. The end points of the open string attached on the D-bane
become non-commutative and consequently the low energy sector of the open string is described 
by the non-commutative field theories \cite{seib}. It is the canonical quantization which shows the
emrgence of the non-commutative algebra, in the most transparent way \cite{Lee;open,lee0105}.   
If both magnetic field and the periodic potential turned on, the model has a more complex 
phase space structure. The non-commutativity and the criticality interplays each other. 
It is noteworthy that the critical behaviour of the Schmid model has been studied in the framework of 
closed string theory while the non-commutativity for the model in the presence the NS B-field has been discussed mostly in the framework of open string theory. It is certainly desirable to discuss both 
features in one framework to understand the complete structure of the dissipative Hofstadter model.   
This paper may be considered as a stepping stone toward the end of a complete analysis of the 
dissipative Hofstadter model. It is shown that the target space duality transformation, $O(2,2;R)$,
removes the interaction with a uniform magnetic field. The $O(2,2;R)$ dual transformation changes the 
basis of oscillator operators so that the algebra of the string coordinate operators at the boundary 
become non-commutative. In the zero temperature limit, the non-commutative algebra of closed string 
coincides with that of open string. It may be also considered as a consequence of the open and closed
string duality. In the open string theory the algebra of the coordinate operators, defined 
at equal $\t$ at end points is non-commutative. In closed string theory, as the world sheet parameters
are interchanged, these points are on the boundary $\t=0$ at equal $\s$. Thus, the non-commutative
algebra of open string must emerge in the low temperature limit or the equal $\s$ limit in the 
closed string theory. 

Immediate applications of this work may be found in condensed matter physics,
since the dissipative Hofstadter model has a wide range of applications as aforementioned.
String theory, which is often referred as a theory of everything, may turn out to be an indispensible tool
to study realistic nanoscale physics as well.

\vskip 1cm

\begin{acknowledgments}
This work was supported by Kangwon National University.
\end{acknowledgments}


%

%

%





\end{document}